%% file: gipsy-type-system-arXiv.tex
\documentclass{easychair}

\pdfoutput=1

\input{commands}

\begin{document}

%
% Title
%

\title{A Type System Theory for Higher-Order Intensional Logic Support for Variable Bindings in Hybrid Intensional-Imperative Programs in {\gipsy}}

\titlerunning{A Type System for HOIL Support in {\gipsy}}

\author{
	{Serguei A. Mokhov}\\
	{Concordia University, Montreal, Canada}\\
	{\url{mokhov@cse.concordia.ca}}\\
\and
	{Joey Paquet}\\
	{Concordia University, Montreal, Canada}\\
	{\url{paquet@cse.concordia.ca}}
}

\authorrunning{Mokhov and Paquet}

\maketitle

%
% Abstract
%

\begin{abstract}
We describe a type system for a platform called the General Intensional Programming System ({\gipsy}), designed to support
intensional programming languages built upon intensional logic and their imperative counter-parts for the intensional execution model.
In {\gipsy}, the type system glues the static and dynamic typing between intensional and imperative
languages in its compiler and run-time environments to support the intensional evaluation of expressions written
in various dialects of the intensional programming language Lucid. The intensionality makes expressions to explicitly take into the account
a multidimensional context of evaluation with the context being a first-class value that serves a number
of applications that need the notion of context to proceed.
We describe and discuss the properties of such a type system and the related type
theory as well as particularities of the semantics, design and implementation
of the GIPSY type system.
\end{abstract}

\section{Introduction}

The {\gipsy} is an ongoing effort for the development of a flexible and adaptable multi-lingual programming language development framework aimed at the investigation on the Lucid family of intensional programming languages~\cite{lucid77,lucid77,lucid85,lucid95,paquetThesis,luthid,eager-translucid-secasa08,multithreaded-translucid-secasa08}.
Using this platform, programs written in many flavors of Lucid can be compiled and executed.
The framework approach adopted is aimed at providing the possibility of easily developing compiler components for other languages of intensional nature, and to execute them on a language-independent run-time system.
{\lucid} being a functional ``data-flow'' language, its programs can be executed in a distributed processing environment.
However, the standard Lucid algebra (i.e. types and operators) is extremely fine-grained and can hardly benefit from parallel evaluation of their operands. Adding granularity to the data elements manipulated by {\lucid} inevitably comes through the addition of
coarser-grained data types and their corresponding operators. Lucid semantics being defined as typeless, a solution to this problem consists in adding a hybrid counterpart to Lucid to allow an external language to define an algebra of coarser-grained types and operators. In turn, this solution raises the need for an elaborated type system to bridge the implicit types of Lucid semantics with the explicit types and operators (i.e. functions) defined in the hybrid counterpart language. This paper presents such a type system that is used in the {\gipsy} at compile time for static type checking, as well as at run-time for dynamic type checking.   

After our problem statement and short presentation of our proposed solution, this section presents a brief introduction to the {\gipsy} project~\cite{gipsy-arch-2000,bolu04,wuf04,dmf-plc05,mokhovmcthesis05,wu05,gipsy2005}, and the beginnings of the GIPSY Type System~\cite{mokhovmcthesis05} in the {\gipsy} to support hybrid and object-oriented intensional programming~\cite{mokhovolucid2005,gipsy-jooip-07}, and {\lucx}'s context type extension known as context calculus~\cite{gipsy-context-calculus-07,gipsy-simple-context-calculus-08,tongxinmcthesis08} for contexts to act as first-class values. Then, we present the complete GIPSY type system as used by the compiler (General Intensional Programming Compiler -- {\gipc}) and the run-time system (General Eduction Engine -- {\gee}) when producing and executing a binary GIPSY program (called General Eduction Engine Resources -- {\geer}) respectively.

\subsection{Problem Statement and Proposed Solution}

\paragraph{Problem Statement}

Data types are implicit in {\lucid} (as well as in its dialects or many functional languages). As such, the type declarations never appear in the Lucid programs at the syntactical level. The data type of a value is inferred from the result of evaluation of an expression. In most imperative languages, like {\java}, {\cpp} and the like, the types are explicit and the programmers must {\em declare} the types of the variables, function parameters and return values before they are used in evaluation. In the {\gipsy}, we want to allow any Lucid dialect to be able to uniformly invoke functions/methods written in imperative languages and the other way around and perform semantic analysis in the form of type assignment and checking statically at compile time or dynamically at run time, perform any type conversion if needed, and evaluate such hybrid expressions. At the same time, we need to allow a programmer to specify, or declare, the types of variables, parameters, and return values for both intensional and imperative functions as a binding contract between inter-language invocations despite the fact that {\lucid} is not explicitly typed. Thus, we need a general type system, well designed and implemented to support such scenarios.

\paragraph{Proposed Solution}

The uniqueness of the type system presented here is that it is above a specific programming language model of either the Lucid family of languages or imperative languages. It is designed to bridge programming language paradigms, the two most general of them would be the intensional and imperative paradigms. {\gipsy} has a framework designed to support a common run-time environment and co-existence of the intensional and imperative languages. Thus, the type system is that of a generic GIPSY program that can include code segments written in a theoretically arbitrary number of intensional and imperative dialects supported by the system vs. being a type system for a specific language. What follows is the details of the proposed solution and the specification of the type system.

\subsection{Introduction to the GIPSY Type System}
\label{sect:gipsy-types}
\index{GIPSY Type System}
\index{GIPSY!Types}
\index{Types}
\index{Frameworks!GIPSY Type System}

The introduction of {\jlucid}, {\olucid}, and the General Imperative Compiler Framework
({\gicf})~\cite{mokhovmcthesis05,mokhovolucid2005,mokhovgicf2005,mokhovjlucid2005}
prompted the development of the GIPSY Type System\index{GIPSY!Type System} as
implicitly understood by the {\lucid} language and its incarnation within
the {\gipsy} to handle types in a more general manner as a glue between
the imperative and intensional languages within the system.
Further evolution of different Lucid variants such as {\lucx} introducing contexts as first-class values
and {\jooip} (Java Object-Oriented Intensional Programming) highlighted the need of the further development of the
type system to accommodate the more general properties of the intensional
and hybrid languages.

\subsubsection{Matching Lucid and Java Data Types}
\label{sect:datatypes-matching}
\index{data types!matching Lucid and Java}

Here we present a case of interaction between {\lucid} and {\java}.
Allowing {\lucid} to call Java methods brings a set of issues related to
the data types, especially when it comes to type checks between Lucid and Java
parts of a hybrid program. This is pertinent when Lucid
variables or expressions are used as parameters to Java methods and when
a Java method returns a result to be assigned to a Lucid
variable or used in an intensional expression. The sets of types in both cases are not exactly
the same. The basic set of Lucid data types as defined by Grogono~\cite{gipcincrements}
is \api{int}, \api{bool}, \api{double}, \api{string},
and \api{dimension}. {\lucid}'s \api{int} is of the same size
as Java's \api{long}. GIPSY and {\java} \api{double}, \api{boolean}, and \api{String} are equivalent.
Lucid \api{string} and Java \api{String} are simply mapped internally through \api{StringBuffer};
thus, one can think of the Lucid \api{string}
as a reference when evaluated in the intensional program. Based on this fact, the lengths
of a Lucid \api{string} and Java \api{String} are the same. Java \api{String} is also
an object in {\java}; however, at this point, a Lucid
program has no direct access to any \api{String}'s properties (though internally we
do and we may expose it later to the programmers).
We also distinguish the \api{float} data type for single-precision floating point operations.
The \api{dimension} index type is said to be an integer or string (as far as its dimension tag values
are concerned), but might be of other types eventually, as discussed in~\cite{gipsy-context-calculus-07}. 
Therefore, we perform data type matching as presented in~\xt{tab:datatypes}.
Additionally, we allow \api{void} Java return type
which will always be matched to a Boolean expression \api{true} in {\lucid} as
an expression has to always evaluate to something.
As for now our types mapping and restrictions are as per~\xt{tab:datatypes}. This is the mapping table for the Java-to-IPL-to-Java type adapter. Such a table would exist for mapping between any imperative-to-intensional language and back, e.g. the {\cpp}-to-IPL-to-{\cpp} type adapter.

\begin{table}
\centering
\caption{Matching data types between Lucid and Java.}
\label{tab:datatypes}
\begin{minipage}[b]{\textwidth}
\begin{center}
\begin{tabular}{|c|c|l|} \hline
{\scriptsize Return Types of Java Methods}  & {\scriptsize Types of Lucid Expressions} & {\scriptsize Internal GIPSY Types}\\ \hline\hline
\api{int}, \api{byte}, \api{long}           & \api{int}, \api{dimension}               & \api{GIPSYInteger} \\
\api{float}                                 & \api{float}                              & \api{GIPSYFloat} \\
\api{double}                                & \api{double}                             & \api{GIPSYDouble} \\
\api{boolean}                               & \api{bool}                               & \api{GIPSYBoolean} \\
\api{char}                                  & \api{char}                               & \api{GIPSYCharacter} \\
\api{String}                                & \api{string}, \api{dimension}            & \api{GIPSYString} \\
\api{Method}                                & {\it function}                           & \api{GIPSYFunction} \\
\api{Method}                                & {\it operator}                           & \api{GIPSYOperator} \\
\api{[]}                                    & \api{[]}                                 & \api{GIPSYArray} \\
\api{Object}                                & {\it class}                              & \api{GIPSYObject} \\
\api{Object}                                & {\it URL}                                & \api{GIPSYEmbed} \\
\api{void}                                  & \api{bool::true}                         & \api{GIPSYVoid} \\ \hline\hline

{\scriptsize Parameter Types Used in Lucid} & {\scriptsize Corresponding Java Types}   & {\scriptsize Internal GIPSY Types} \\ \hline\hline
\api{string}                                & \api{String}                             & \api{GIPSYString}\\
\api{float}                                 & \api{float}                              & \api{GIPSYFloat}\\
\api{double}                                & \api{double}                             & \api{GIPSYDouble}\\
\api{int}                                   & \api{int}                                & \api{GIPSYInteger} \\
\api{dimension}                             & \api{int}, \api{String}                  & \api{Dimension} \\
\api{bool}                                  & \api{boolean}                            & \api{GIPSYBoolean}\\
{\it class}                                 & \api{Object}                             & \api{GIPSYObject}\\
{\it URL}                                   & \api{Object}                             & \api{GIPSYEmbed}\\
\api{[]}                                    & \api{[]}                                 & \api{GIPSYArray}\\
{\it operator}                              & \api{Method}                             & \api{GIPSYOperator}\\
{\it function}                              & \api{Method}                             & \api{GIPSYFunction}\\ \hline
\end{tabular}
\end{center}
\end{minipage}
\end{table}

\section{Simple Theory of GIPSY Types}
\label{sect:stgt}

Our simple theory of the GIPSY types (STGT) is based on the
``Simple theory of types'' (STT) by Mendelson~\cite{math-logic-97}.
The theoretical and practical considerations are described in the sections
that follows. The STT partitions the qualification domain into
an ascending hierarchy of types with every individual
value assigned a type. The type assignment is dynamic
for the intensional dialects as the resulting type of
a value in an intensional expression may not be known
at compile time. The assignment of the types of constant literals
is done at compile-time, however. In the hybrid system,
which is mostly statically typed at the code-segment boundaries, the type assignment
also occurs at compile-time. On the boundary between the
intensional programming languages ({\ipl}s) and imperative languages, prior to an imperative procedure
being invoked, the type assignment to the procedure's parameters
from IPL's expression is computed dynamically {\em and} matched
against a type mapping table similar to that of \xt{tab:datatypes}.
Subsequently, the when the procedure call returns back to the IPL,
the type of the imperative expression is matched back through
the table and assigned to the intensional expression that expects it.
The same table is used when the call is made by the procedure
to the IPL and back, but in the reverse order.

Further in STT, all quantified
variables range over only one type making the first-order logic
applicable as the underlying logic for the theory. This also means
the all elements in the domain and all co-domains are of the same
type. The STT states there is an atomic type that has no member
elements in it, and the members of the second-high from the
basic atomic type. Each type has a next higher type similarly
to \api{succ} in Peano arithmetic and the \api{next} operator
in {\lucid}. This is also consistent to describe the composite
types, such as arrays and objects as they can be recursively
decomposed (or ``flattened'', see~\cite{mokhovmcthesis05}) into the primitive
types to which the STT applies.

Symbolically, the STT uses primed and unprimed variables and the
infix set notation of $\in$. The formulas $\Phi(x)$ rely on the
fact that the unprimed variables are all of the same type. This
is similar to the notion of a Lucid stream with the point-wise
elements of the stream having the same type. The primed variables ($x'$)
in STT range over the next higher type. There two atomic formulas
in STT of the form of identity, $x=y$, and set membership, $y \in x'$.

The STT defines the four basic axioms for the variables and the
types they can range over: Identity, Extensionality, Comprehension,
and Infinity. We add the Intensionality on as the fifth axiom. The variables in the definition
of the Identity relationship and in the Extensionality and Comprehension
axioms typically range over the elements of one of the two
nearby types. In the set membership, only the unprimed variables
that range over the lower type in the hierarchy can appear on the
left of $\in$; conversely, the primed ones that range over higher
types can only appear on the right of $\in$. The axioms are defined as:

\begin{enumerate}
\item
{\bf Identity}: $x = y \leftrightarrow \forall z' [x \in z' \leftrightarrow y \in z']$
\item
{\bf Extensionality}: $\forall x[x \in y' \leftrightarrow x \in z'] \rightarrow y' = z'$
\item
{\bf Comprehension}: $\exists z' \forall x[x \in z' \leftrightarrow \Phi(x)]$. This covers objects and arrays as any collection of elements here may form
an object of the next, higher type. The STT states the comprehension axiom is schematic with respect to $\Phi(x)$ and the types.
\item
{\bf Infinity}: $\forall x,y[x \neq y \rightarrow [xRy \bigvee yRx]]$. There exists a non-empty binary relation $R$ over the elements of the atomic type that is transitive, irreflexive, and strongly connected. 
\item
{\bf Intensionality}: the intensional types and operators are based on the intensional logic and context calculus. These are extensively described in the works~\cite{gipsy-simple-context-calculus-08,gipsy-context-calculus-07,wanphd06,paquetThesis} and related. This present type system accommodates the two in a common
hybrid execution environment of the {\gipsy}.
A context $c$ is a finite subset of the relation: $c \subset \{(d, x) | d \in DIM \wedge x \in T\}$, where $DIM$ is the set of all possible dimensions, and $T$ is the set of all possible tags.
\end{enumerate}

\subsection{Types of Types}

Types of types are generally referred to as {\em kinds}.
Kinds provide categorization to the types of similar nature.
While some type systems provide kinds as first class entities
available to programmers, in {\gipsy} we do not expose this
functionality in our type system at this point. However,
at the implementation level there are provisions to do so
that we may later decide to expose for the use of programmers.
Internally, we define several broad kinds of types, presented
the the sections that follow.

\subsubsection{Numeric Kind}
\label{sect:numeric-types}

The primitive types under this category are numerical values,
which are represented by \api{GIPSYInteger},
\api{GIPSYFloat}, and \api{GIPSYDouble}.
They provide implementation of the common arithmetic operators,
such as addition, multiplication and so on, as well as logical
comparison operators of ordering and equality. Thus, for a
numerical type $T$, the following common operators are provided.
The resulting type of any arithmetic operator is the largest
of the two operands in terms of length (the range of \api{double}
of length say $k$
covers the range of \api{int} with the length say $m$ and if both appear as arguments
to the operator, then the resulting value's type is that of
without loss of information, i.e. largest in length \api{double}).
The result of the logical comparison operators is always Boolean $B$
regardless the length of the left-hand-side and right-hand-side
numerical types.

% \vspace{3mm}
\begin{enumerate}
\item $T_{max}: T_1 \geq T_2 \rightarrow T_1 \;|\; T_1 < T_2 \rightarrow T_2$
\item $T_{multiply}: T_k \times T_m \rightarrow T_{\max({k,m})}$
\item $T_{divide}: T_k / T_m \rightarrow T_{\max({k,m})}$
\item $T_{add}: T_k + T_m \rightarrow T_{\max({k,m})}$
\item $T_{subtract}: T_k - T_m \rightarrow T_{\max({k,m})}$
\item $T_{mod}: T_k \% T_m \rightarrow T_{\max({k,m})}$
\item $T_{pow}: T_{k}\; \hat\;\; T_{m} \rightarrow T_{\max({k,m})}$
\item $T_{>}: T > T \rightarrow B$
\item $T_{<}: T < T \rightarrow B$
\item $T_{\geq}: T \geq T \rightarrow B$
\item $T_{\leq}: T \leq T \rightarrow B$
\item $T_{=}: T = T \rightarrow B$
\end{enumerate}
% \vspace{3mm}

A generalized implementation of the arithmetic operators is done by realization of the interface
called \api{IArithmeticOperatorsProvider} and its concrete
implementation developed in the general delegate class \api{GenericArithmeticOperatorsDelegate}.
This design and implementation allow not only further
exposure of kinds-as-first-class values later on after
several iterations of refinement, but also will allow
operator and type overloading or replacement of type handling
implementation altogether if some researchers wish to
do so. The implementation of the engine, {\gee}, is thus
changed, to only refer to the interface type implementation
when dealing with these operators. Equivalently for the logic
comparison operators we have the \api{ILogicComparisonOperatorsProvider} class
and the corresponding  \api{GenericLogicComparisonOperatorsDelegate} class. The latter
relies on the comparator implemented for the numerical kind,
such as \api{NumericComparator}. Using comparators (i.e. classes
that implement the standard \api{Comparator} interface) allows
{\java} to use and to optimize build-in sorting and searching
algorithms for collections of user-defined types. In our case,
the class called \api{GenericLogicComparisonOperatorsDelegate} is the implementation of the delegate class
that also relies on it. The example for the numeric types for the
described design is in \xf{fig:providers-comparators}.

\begin{figure*}[htp!]
	\begin{centering}
	\includegraphics[width=.7\textwidth]{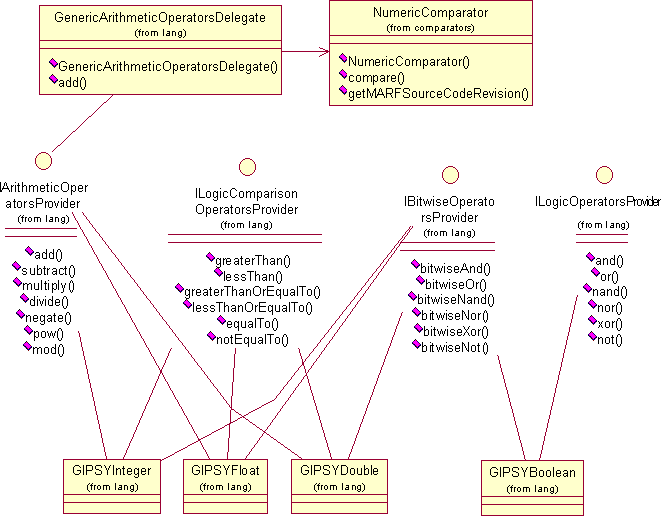}
	\caption{Example of Provider Interfaces and Comparators}
	\label{fig:providers-comparators}
	\end{centering}
\end{figure*}

It is important to mention, that grouping of the numeric
kind of integers and floating-point numbers does not violate
the IEEE~754 standard~\cite{ieee754}, as these kinds implementation-wise
wrap the corresponding {\java}'s types (which are also
grouped under numeric kind) and their semantics including
the implementation of IEEE~754 by {\java} in accordance
with the Java Language Specification.

\subsubsection{Logic Kind}
\label{sect:logic-types}

Similarly to numeric types, the primitive type
\api{GIPSYBoolean} fits the category of the types
that can be used in the Boolean expressions. The operations
the type provides expect the arguments to be of the same
type~--~Boolean. The following set of operators on the
logic type $B$ we provide in the GIPSY type system:

% \vspace{3mm}
\begin{enumerate}
\item $B_{and}: B \bigwedge B \rightarrow B$
\item $B_{or}: B \bigvee B \rightarrow B$
\item $B_{not}: \neg B \rightarrow B$
\item $B_{xor}: B \bigoplus B \rightarrow B$
\item $B_{nand}: \neg (B \bigwedge B) \rightarrow B$
\item $B_{nor}: \neg (B \bigvee B) \rightarrow B$
\end{enumerate}
% \vspace{3mm}

\noindent
Note that the logical XOR operator (denoted as $\bigoplus$) is different
from the corresponding bitwise operator in \xs{sect:bitwise-types}
similarly to as the bitwise vs. logical are for AND and OR. Again, similarly to the generalized implementation of arithmetic
operators, logic operator providers implement the \api{ILogicOperatorsProvider}
interface, with the most general implementation of it in the delegate class \api{GenericLogicOperatorsDelegate}.

\subsubsection{Bitwise Kind}
\label{sect:bitwise-types}

Bitwise kind of types covers all the types that can support
bitwise operations on the the entire bit length of a particular
type $T$. Types in this category include the numerical and
logic kinds described earlier in \xs{sect:logic-types} and
\xs{sect:numeric-types}. The parameters on both sides of the
operators and the resulting type are always the same. There
are no implicit compatible type casts performed unlike for
the numeric kind.

% \vspace{3mm}
\begin{enumerate}
\item $T_{bit-and}: T \& T \rightarrow T$
\item $T_{bit-or}: T | T \rightarrow T$
\item $T_{bit-not}: \;! T \rightarrow T$
\item $T_{bit-xor}: T \bigotimes T \rightarrow T$
\item $T_{bit-nand}: \;! (T \& T) \rightarrow T$
\item $T_{bit-nor}: \;! (T | T) \rightarrow T$
\end{enumerate}
% \vspace{3mm}

\noindent
The generalized implementation of this kind's operators is done through
the interface that we called \api{IBitwiseOperatorsProvider}, which is generically implemented
in the corresponding delegate class \api{GenericBitwiseOperatorsDelegate}.

\subsubsection{Composite Kind}

As the name suggests, the composite kind types consist of compositions of other types,
possibly basic types. The examples of this kind are arrays, structs,
and abstract data types and their realization such as objects and collections.
In the studied type system these are \api{GIPSYObject}, \api{GIPSYArray},
and \api{GIPSYEmbed}. This kind is characterized by the constructors, dot operator
to decide membership as well as to invoke member methods and define
equality. The design of of these types, just like the entire type system,
adheres to the Composite design pattern. The most prominent examples of
this kind are in \xf{fig:gipsy-types-composite}, including \api{GIPSYContext}
which is composed of \api{Dimension}s and indirectly of the \api{TagSet}s.

\begin{figure*}[htp!]
	\begin{centering}
	\includegraphics[width=.8\textwidth]{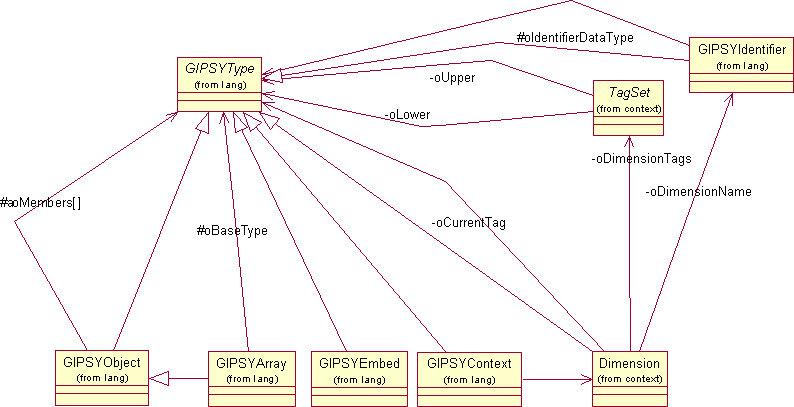}
	\caption{Composite Types.}
	\label{fig:gipsy-types-composite}
	\end{centering}
\end{figure*}

\subsubsection{Intensional Kind}

The intentional types kind primarily has to do with encapsulation
dimensionality, context information and their operators.
These are represented by the types \api{GIPSYContext}, \api{Dimension},
and \api{TagSet}\footnote{Not mentioned here; please refer
to~\cite{tongxinmcthesis08,gipsy-simple-context-calculus-08}.} types. The common operators
on these types include the context switching and querying operators
@ and \# as well as context calculus operators. Additional
operators can be included depending on the intensional dialect used,
but the mentioned operators are said to be the baseline operators
that any intensional language can be translated to use.
Implementation-wise there is a \api{IContextSetOperatorsProvider}
interface and its general implementation in \api{GenericContextSetOperatorsDelegate}.
The context calculus operators on simple contexts
include standard set operators, such as {$\mathbf{union}$}, {$\mathbf{difference}$}, {$\mathbf{intersection}$}, %\api{subset},
and {\lucx}-specific {$\mathbf{isSubContext}$}, {$\mathbf{projection}$}, {$\mathbf{hiding}$},
and {$\mathbf{override}$}~\cite{tongxinmcthesis08,gipsy-simple-context-calculus-08}.

\subsubsection{Function Kind}

The function types represent either ``functional'' functions, imperative
procedures, and binary and unary operators that, themselves, can be
passed as parameters or returned as results. In our type system
these are represented by \api{GIPSYFunction}, \api{GIPSYOperator},
and \api{GIPSYEmbed}. The common operators on this include equality
and evaluation.

\section{Describing Some GIPSY Types' Properties}
\label{sect:gipsy-types-properties}
\label{appdx:gipsy-types-properties}

To show the overview of the properties of GIPSY types
to get a better and more complete understanding of the spectrum of
their behavior we cover them in light of description
and comparison to existential, union, intersection,
and linear types.

\subsection{Existential Types}

All of the presented GIPSY types are existential types because
they represent and encapsulate modules and abstract data types and
separate their implementation from the public interface specified
by the abstract \api{GIPSYType}. These are defined as follows:

\vspace{3mm}
{\small{$T = \exists \mathtt{GIPSYType} \{ \mathtt{Object}\; a;\; \mathtt{Object\; getEnclosedTypeObject()};\; \mathtt{Object\; getValue()};\}$}}
\vspace{0mm}

\noindent
The above is implemented in the following manner, e.g.:

\vspace{3mm}
\begin{small}
$boolT\; \{ \mathtt{Boolean}\; a; \mathtt{Object\; getEnclosedTypeObject()};\; \mathtt{Boolean\; getValue()}; \}$

$intT\; \{ \mathtt{Long}\; a; \mathtt{Object\; getEnclosedTypeObject()};\; \mathtt{Long\; getValue()}; \}$

$doubleT\; \{ \mathtt{Double}\; a; \mathtt{Object\; getEnclosedTypeObject()};\; \mathtt{Double\; getValue()}; \}$

%$functionT\; \{ \mathtt{FunctionItem}\; a; \mathtt{Object\; getEnclosedTypeObject()};\; \mathtt{FunctionItem\; getValue()}; \}$

... and so on.
\end{small}
\vspace{3mm}

All these correspond to be subtypes of the the more abstract and general existential type $T$. Assuming the value $t \in T$, then $t.\mathtt{getEnclosedTypeObject()}$ and $t.\mathtt{getValue()}$ are well typed regardless the what the \api{GIPSYType} may be.

\subsection{Union Types}

A union of two types produces another type with the valid range of
values that is valid in {\em either} of the two; however, the operators
defined on the union types must be those that are valid for {\em both}
of the types to remain type-safe. A classical example of that is in {\C} or {\cpp}
the range for the \api{signed char} is $-128 \ldots 127$ and
the range of the \api{unsigned char} is $0 \ldots 255$, thus:

\vspace{3mm}
$\mathtt{signed\; char} \cup \mathtt{unsigned\; char} = -128 \ldots 255$
\vspace{3mm}

\noindent
In {\C} and {\cpp} there is a \api{union} type that roughly corresponds
to this notion, but it does not enforce the operations that are possible
on the union type that must be possible in both left and right types of
the uniting operator. In the class hierarchy, such as in {\gipsy}, the union type among the
type and its parent is the parent class; thus, in our specific type
system the following holds:

\vspace{3mm}

$\forall T \in G: T \cup \mathtt{GIPSYType} = \mathtt{GIPSYType}$

$\mathtt{GIPSYArray} \cup \mathtt{GIPSYObject} = \mathtt{GIPSYObject}$

$\mathtt{GIPSYVoid} \cup \mathtt{GIPSYBoolean} = \mathtt{GIPSYBoolean}$

$\mathtt{GIPSYOperator} \cup \mathtt{GIPSYFunction} = \mathtt{GIPSYFunction}$

\vspace{3mm}

\noindent
where $T$ is any concrete GIPSY type and $G$ is a collection of types
in the GIPSY type system we are describing. Equivalently, the union of the two sibling types is their common
parent class in the inheritance hierarchy. Interestingly enough,
while we do not explicitly expose kinds of types, we still are able
to have the following union type relationships defined based on the
kind of operators they provide as siblings under a common interface:

\vspace{3mm}
\begin{small}
$\forall T \in A: T \cup \mathtt{IArithmeticOperatorProvider} = \mathtt{IArithmeticOperatorProvider}$

$\forall T \in L: T \cup \mathtt{ILogicOperatorProvider} = \mathtt{ILogicOperatorProvider}$

$\forall T \in B: T \cup \mathtt{IBitwiseOperatorProvider} = \mathtt{IBitwiseOperatorProvider}$

$\forall T \in C: T \cup \mathtt{IContextOperatorProvider} = \mathtt{IContextOperatorProvider}$

$\forall T \in D: T \cup \mathtt{ICompositeOperatorProvider} = \mathtt{ICompositeOperatorProvider}$

$\forall T \in F: T \cup \mathtt{IFunctionOperatorProvider} = \mathtt{IFunctionOperatorProvider}$
\end{small}
\vspace{3mm}

\noindent
where $T$ is any concrete GIPSY type and $A$ is a collection of types
that provide arithmetic operators, $L$ -- logic operators providers, $B$ -- bitwise
operators providers,  $C$ -- context operators providers, $D$ -- composite operator providers,
and $F$ -- function operator providers. Thus:

\vspace{3mm}
\begin{small}
$\{\mathtt{GIPSYInteger}, \mathtt{GIPSYFloat}, \mathtt{GIPSYDouble}\} \in A$

$\{\mathtt{GIPSYBoolean}\} \in L$

$\{\mathtt{GIPSYInteger}, \mathtt{GIPSYFloat}, \mathtt{GIPSYDouble}, \mathtt{GIPSYBoolean}\} \in B$

$\{\mathtt{GIPSYContext}, \mathtt{Dimension}\} \in C$

$\{\mathtt{GIPSYObject}, \mathtt{GIPSYArray}, \mathtt{GIPSYEmbed}\}, \mathtt{GIPSYString}\} \in D$

$\{\mathtt{GIPSYFunction}, \mathtt{GIPSYOperator}, \mathtt{GIPSYEmbed}\} \in F$
\end{small}
\vspace{3mm}

\noindent
Another particularity of the GIPSY type system is the union of
the string and integer types under dimension:

\vspace{3mm}
$\mathtt{GIPSYInteger} \cup \mathtt{GIPSYString} = \mathtt{Dimension}$
\vspace{3mm}

\noindent
and this is because in our dimension tag values we allow them to be
either integers or strings. While not a very common union in the
majority of type system, they do share a common set of tag set
operators defined in \cite{gipsy-context-calculus-07} for ordered
finite tag sets (e.g. \api{next()}, etc.).

\subsection{Intersection Types}

Intersection type of given two types is a range where the sets
of valid values overlap. Such types are safe to pass to methods
and functions that expect either of the types as the intersection
types are more restrictive and compatible in both original types.
A classical example of an intersection
type if it were implemented in {\C} or {\cpp} would be:

\vspace{3mm}
$\mathtt{signed\; char} \cap \mathtt{unsigned\; char} = 0 \ldots 127$
\vspace{3mm}

\noindent The intersection types are also useful in describing the overloaded
functions. Sometimes the are called as refinement types. In the class
hierarchy, the intersection between the
parent and child classes is the most derived type, and the intersection
of the sibling classes is empty. While the functionality offered
by the intersection types is promising, it is not currently
explicitly or implicitly considered in the {\gipsy} type system,
but planned for the future work.

\subsection{Linear Types}

Linear (or ``uniqueness'') types are based on linear logic. The main
idea of these types is that values assigned to them have one and only one
reference to them throughout. These types are useful to describe
immutable values like strings or hybrid
intensional-imperative objects (see~\cite{gipsy-jooip-07} for details).
These are useful because most operations on such an object ``destroys'' it
and creates a similar object with the new values, and, therefore,
can be optimized in the implementation for the in-place mutation.
Implicit examples of such types in the GIPSY type system are \api{GIPSYString}
that internally relies on {\java}'s \api{StringBuffer} that does something
very similar as well as the immutable \api{GIPSYObject} is in {\jooip} \cite{gipsy-jooip-07}
and immutable \api{GIPSYFunction}. Since we either copy or retain the values in the warehouse, and, therefore,
one does not violate referential transparency or create side effects in, and at the same time be more efficient
as there is no need to worry about synchronization overhead.

\section{Conclusion}

Through a series of discussions, specification, design, and implementation
details we presented a type system used by the {\gipsy} for static and
dynamic type checking and evaluation of intensional and hybrid intensional-imperative
languages. We highlighted the particularities of the system that does not
attribute to a particular specific language as traditionally most type
systems do, but to an entire set of languages and hybrid paradigms that
are linked through the type system. We pointed out some of the
limitations of the type system and the projected work to remedy
those limitations. Overall, this is a necessary contribution to
GIPSY-like systems to have a homogeneous environment to statically
and dynamically type-check intensional and hybrid programs.

\subsection{Future Work}

The type system described in this paper has been implemented in the GIPSY.
However, due to recent changes including the introduction of contexts as first-class
values in {\em Generic Lucid}, as well as the development of the fully-integrated
hybrid language JOOIP~\cite{gipsy-jooip-07}, our implementation still needs
thorough testing using complex program examples testing the limits of the
type system. Additional endeavours, noted in the previous sections of this paper, include: 

\begin{itemize}
\item
Exposing more operators for composite types to Lucid code segments.

\item
Allowing custom user-defined types and extension of the existing operators and operator overloading.

\item
Expose {\em kinds} as first class entities, allowing programs more explicit manipulation of types. 

\item
Consider adding intersection types for more flexibility in the future development of the type system, allowing more type casting possibilities at the programming level. 

\end{itemize}

\subsection{Acknowledgments}

This work was funded by NSERC and the Faculty of Computer Science and Engineering, Concordia University.

%%%%%%%%%%%%%%%%%%%%%%%%%%%%%%%%%%%%%%
% Refs:
%
\label{sect:bib}
\bibliographystyle{abbrv}
\bibliography{gipsy-type-system-arXiv}

\end{document}

%% file: commands.tex
\newcommand{\xf}[1]{Figure~\ref{#1}}

\newcommand{\xs}[1]{Section~\ref{#1}}

\newcommand{\xt}[1]{Table~\ref{#1}}

\newcommand{\gipc}{{GIPC\index{GIPC}\index{Frameworks!GIPC}}}
\newcommand{\gicf}{{GICF\index{GICF}\index{Frameworks!GICF}}}

\newcommand{\gee}{{GEE\index{GEE}\index{Frameworks!GEE}}}
\newcommand{\geer}{{GEER\index{GEER}}}
\newcommand{\gipsy}{{GIPSY\index{GIPSY}}}

\newcommand{\ipl}{{IPL\index{IPL}}}
\newcommand{\lucid}{{Lucid\index{Lucid}}}

\newcommand{\jlucid}{{JLucid\index{JLucid}}}
\newcommand{\olucid}{{Objective Lucid\index{Tensor Lucid}}}

\newcommand{\lucx}{{Lucx\index{Lucx}}}

\newcommand{\jooip}{{JOOIP\index{JOOIP}}}

\newcommand{\C}{{C\index{C}}}
\newcommand{\cpp}{{C++\index{C++}}}

\newcommand{\java}{{Java\index{Java}}}

\newcommand{\api}[1]{\texttt{#1}\index{API!#1}}

\newcommand{\lucidL}[1]{{$\mathit{Lucid}$}($L$) }

		{}

\def\myvert{\raise 2.27pt \hbox{\vrule depth 0pt height 8pt width 0.2mm}}
\def\myarrow{\hspace*{0.43mm}%
             \raise 2.29pt\hbox{\vrule depth 0pt height 8pt width 0.16mm}%
             \hspace*{-0.32mm}%
             $\longrightarrow$
             \ %
             }